\documentclass[letterpaper,11pt]{article}
\usepackage{jheppub_mod}
\usepackage{graphicx, amsmath, amssymb, amstext}
\usepackage{relsize, float, multirow}
\usepackage{array, epstopdf, hyperref, }
\usepackage{caption, subcaption, mathtools}
\usepackage{newtxmath}
\usepackage[normalem]{ulem}

\DeclareMathOperator{\erfc}{erfc}

\title{CMB and BBN constraints on evaporating primordial black holes revisited}
\author[a]{Sandeep Kumar Acharya,} 
\author[a]{Rishi Khatri}
\affiliation[a]{Department of Theoretical Physics, Tata Institute of 
Fundamental Research, Mumbai 400005, India}
\emailAdd{sandeepkumar@theory.tifr.res.in, khatri@theory.tifr.res.in}
\date{\today}
\abstract{ 
We derive new CMB anisotropy power spectrum and BBN constraints for
evaporating primordial black holes with mass between $10^{11}$~g and $10^{16}$~g by explicitly solving the
electromagnetic particle cascades of emitted particles and the deposition
of this emitted energy to the background baryon-photon plasma. We show that
the CMB anisotropies can provide stronger constraints compared to
BBN and CMB spectral distortions on black holes with masses as small as
$M_{\rm BH}=1.1\times 10^{13}$g, a slightly smaller mass than what has been
considered in literature until now. We also show that, with more up-to-date
data on abundances of deuterium and helium-3, BBN constraints are
strengthened significantly. The abundance of these primordial black holes
constrains the epoch of inflation $\sim 40$ e-folds after the epoch
constrained by the CMB observations.
}
\notoc
\begin{document}
\maketitle
\newpage
\section{\label{sec:intro}Introduction}
Primordial black holes (PBH) have recently gathered attention as an
explanation of dark matter \cite{BCMAKKRR2016} in light of the discovery of
gravitational waves from the merger of black holes with the mass of tens of
solar mass \cite{LIGO2016}. Regardless of whether PBH form a dominant
component of dark matter today, evidence for their existence at any time in
cosmic history, in any mass range, would give great insight into the
initial conditions and, in particular, the initial density fluctuations of
the universe (for a recent summary of these constraints, see
\cite{CKS2016}). PBHs are formed in the radiation dominated era when the
radiation pressure is not able to resist the gravitational collapse in
overdense regions \cite{ZN1966,H1971,CH1974,NPSZ1979}. Mass of the black hole,
 produced in this scenario, is of the order of horizon mass
\cite{CH1974,CKSY2010}. Due to formation at different epochs, mass of PBHs
can vary from Planck mass relics to $\sim 10^{10}$ times heavier than the
mass of the Sun \cite{CKS2016}. Although PBH formation with extended mass
spectrum is likely \cite{KF2017,CRTVV2017}, we will restrict ourselves to
monochromatic PBH mass function in this work so as not to restrict
ourselves to a particular model. Since PBHs are formed due to early
universe small scale fluctuations, their abundance today puts constraints
on the initial power spectrum on small scales
\cite{C1975,ES2018}. Therefore, deriving accurate constraints on the allowed abundance of PBHs through various cosmological probes is a subject of great interest today. In this paper, we calculate constraints on the allowed abundance of PBHs in the mass range $\sim 10^{11}-10^{17}$g from CMB (Cosmic Microwave Background) anisotropy power spectrum, CMB spectral distortions and BBN (Big bang nucleosynthesis). There are constraints on accreting solar mass black holes  in the early universe from CMB spectral distortions and CMB anisotropy power spectrum \cite{ROM2008,Am2017,PSCCK2017} but we will not consider them here. \par
\hspace{1cm}
Injection of energetic electrons, positrons and photons around
the recombination era (at redshift $z\sim 1000$) can heat the background
baryon-photon plasma or, if the injected particles  are
sufficiently energetic ($>10.2$ eV), they can excite and ionize
neutral hydrogen and helium. Additional ionizations due to extra energy
injections lead to a higher freeze-out free electron fraction at
  the end of the recombination epoch ($z\lesssim 1000$) as compared to the
standard recombination history \cite{Zks1969,Peebles1969}. Increased
scattering of CMB photons with free electrons leads to the damping of the
CMB temperature anisotropy power spectrum and a boost in the polarization
on large scales
\cite{ASS1998,CK2003,Padmanabhan:2005es,Slatyer:2009yq,Galli:2009zc}.
Electromagnetic energy injection at $z\lesssim 2\times 10^6$ into the
baryon-photon fluid results in $y,i$, and $\mu$-type spectral
  distortions (collectively called $yim$ distortions  hereafter) in the non-relativistic
  theory
\cite{Sz1969,Sz19701,Is19752,SC1984,Bdd1991,Chluba:2011hw,Ks2012b,Chluba:2013vsa}.
Recently, it has been shown that for energy injection at $z\lesssim 2\times
10^5$, relativistic effects are important \cite{AK2018,AK2019}. At
$z>2\times 10^6$, photon number non-conserving processes establish a Planck spectrum and erase any CMB spectral distortions \cite{Sz19701,dd1982,Chluba:2011hw,ks2012}. Energetic photons above the  photo-dissociation threshold of deuterium and helium-4 can change the abundances of deuterium, helium-3 and helium-4 \cite{ZSKC1977,ENS1985,EGLNS1992,KM1995,KKMT2018,PS2015,HSW2018,FMW2019}. These energetic photons are efficient in photo-dissociating nuclei upto a redshift at which the pair-production threshold on the CMB photons is of the same order as the photo-dissociation threshold of the nuclei \cite{PS2015,HSW2018,FMW2019}. Above this critical redshift, $z\approx 3\times 10^6$ for helium-4 destruction and $z\approx 3\times 10^7$ for deuterium destruction, constraints become exponentially weak as most of the energy is lost to pair-production on the CMB photons instead of destruction of BBN elements \cite{KM1995,PS2015}.\par
\hspace{1cm}
Black holes emit energetic particles by Hawking evaporation
\cite{H1974,H1975}. The effect of PBH evaporation on the CMB anisotropy
power spectrum was studied in \cite{PLS2017,CDGSW2017,SKLP2018}. The
authors in \cite{PLS2017} only considered energy deposition from primary electrons, positrons and photons and neglected the decay products of heavier particles. Also, they ignored the evolution of energy spectra during black hole evaporation. The
authors of \cite{CDGSW2017} derived constraints for PBH in mass range
$10^{15}-10^{17}$g with 6 cosmological parameters \cite{Pl2018} fixed. The
authors in \cite{SKLP2018} provided CMB anisotropy constraints for
monochromatic PBHs, taking into account secondary photons,
electron-positron pairs after hadronization from quarks and gluons. For the
energy deposition of high energy electromagnetic particles to the
background medium, the authors of \cite{PLS2017,CDGSW2017,SKLP2018} used
the result of \cite{Slatyer20162}. Similar analysis was performed in
\cite{LSHLC2019} but with on-the-spot approximation which means that any
energy injected at a particular redshift is deposited to the background
baryon-photon fluid at that particular redshift.
    The low
optical depth of high energy gamma rays \cite{Slatyer:2009yq} implies that
the energetic photons deposit their energy gradually over an extended
period of time.  The authors in \cite{PAHWW2019} derived constraints for
PBHs in the mass range $10^{15}-10^{17}$g without the on-the-spot
approximation and using the result of \cite{Slatyer20162}. It was recently
shown that some of the energy inject even at $z\sim 10000$ can survive
until recombination and have an observable effect on the CMB anisotropy
power spectrum \cite{AK20192} \par
\hspace{1cm}               
 In this paper, we derive CMB anisotropy constraints using our own recent
 calculations \cite{AK20192}, by evolving the electromagnetic cascades in
 the expanding universe without assuming on-the-spot approximation. We
 extend the CMB anisotropy constraints to lower PBH mass or PBH decaying at
 higher redshifts upto the point at which the CMB anisotropy
 constraints are not competitive with the constraints from CMB spectral distortions and BBN. We find that the CMB anisotropy constraints are competitive for PBH mass $\gtrsim 1.3\times 10^{13}$g, a slightly lower mass threshold than \cite{SKLP2018}. We also provide BBN constraints with a more up-to-date analysis of electromagnetic cascade with Hubble expansion taken into account\cite{AK20192} and obtain stronger constraints compared to  \cite{KY2000,CKSY2010}. Our aim in this paper is to provide more accurate constraints from CMB probes and BBN. We like to point out that other cosmological signals like the  global 21 cm signal \cite{CDGMS2018} and astrophysical probes like gamma rays \cite{CKSY2016,AAS2019}, cosmic rays \cite{BC2019}, 511 keV gamma rays line from galactic center \cite{DG2019,L2019,DLR2019} can give competitive or stronger constraints at the higher end of the mass range constrained by the CMB anisotropies. \par
 \hspace{1cm}
 We use Planck \cite{Pl2018} cosmological parameters in all calculations.
 \section{Energy deposition from electromagnetic cascade}
     A high energy injected particle can boost a background electron or
     photon to a higher energy in a relativistic scattering. This boosted
     particle along with the original particle (with reduced energy
    compared to the original energy) can boost more background
     particles, creating more relativistic particles and thus causing an
     electromagnetic cascade. The average energy of a particle in the
     cascade decreases as more energetic particles are produced. In an
     ionized universe, these particles can only heat the background
     electrons. For a partially neutral medium, sub-keV photons can
     photo-ionize and excite neutral atoms. Electrons and positrons with
     energy $\gtrsim$keV boost the CMB photons through inverse Compton
     scattering (IC). Photons below the Lyman-alpha threshold (=10.2 eV)
     escape. For electrons and positrons, the energy loss rates are much
     faster compared to the Hubble rate
     \cite{Slatyer:2009yq,AK2018}. Therefore, electrons and positrons lose
     their energy instantaneously through IC to the CMB photons or as heat
     to background electrons. The energy loss rate of photons is comparable to the Hubble rate. Therefore, we have to evolve the photon spectrum with all the relevant scattering processes in an expanding universe. For such calculations, we have followed the approach of \cite{Kanzaki:2008qb,Kanzaki:2009hf,Slatyer:2009yq}. Our calculation of energy deposition of electromagnetic particles to background medium  is detailed in \cite{AK20192}. The basic algorithm to evolve the cascades is following. We divide the energy range from $\sim$eV to $\sim$TeV in 200 log spaced energy bins. A particle, in a particular energy bin, after depositing a fraction of its  energy to the background particles can only drop down to lower energy bins. Therefore, the information of subsequent electromagnetic cascade of lower energy particles can be used to calculate to full cascade evolution history of original injected particle. Starting from lowest energy bins with energy of the order of few eV, for which electrons can only heat the background medium and photons which can ionize the neutral atoms, we can compute the electromagnetic cascade and energy deposition fraction of higher energy injected particles. The calculation thus proceeds in a recursive way, from low energy to high energies. More details can be found in \cite{AK20192}.      
     \section{Time evolution of evaporating black holes}
\begin{figure}[!tbp]
  \begin{subfigure}[b]{0.4\textwidth}
    \includegraphics[scale=1.0]{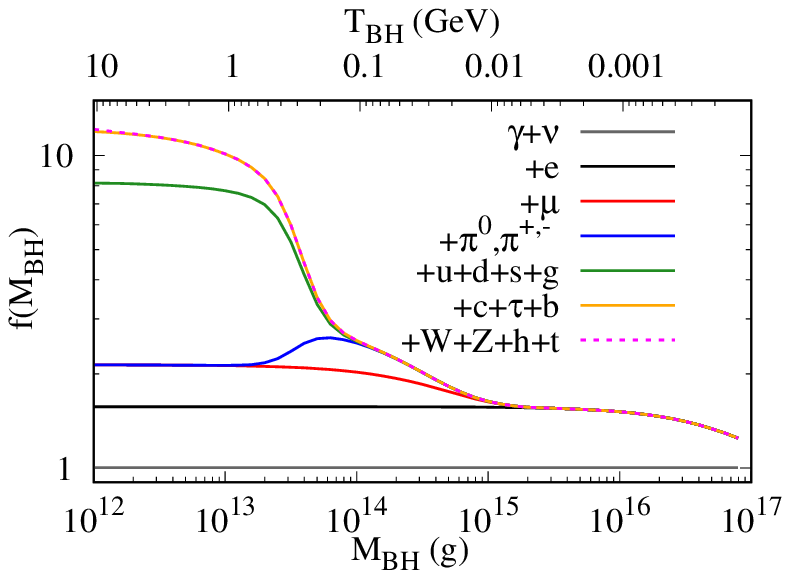}
    \caption{}
    \label{fig:fmbh}
  \end{subfigure}
  \hfill
  \begin{subfigure}[b]{0.4\textwidth}
    \includegraphics[scale=1.0]{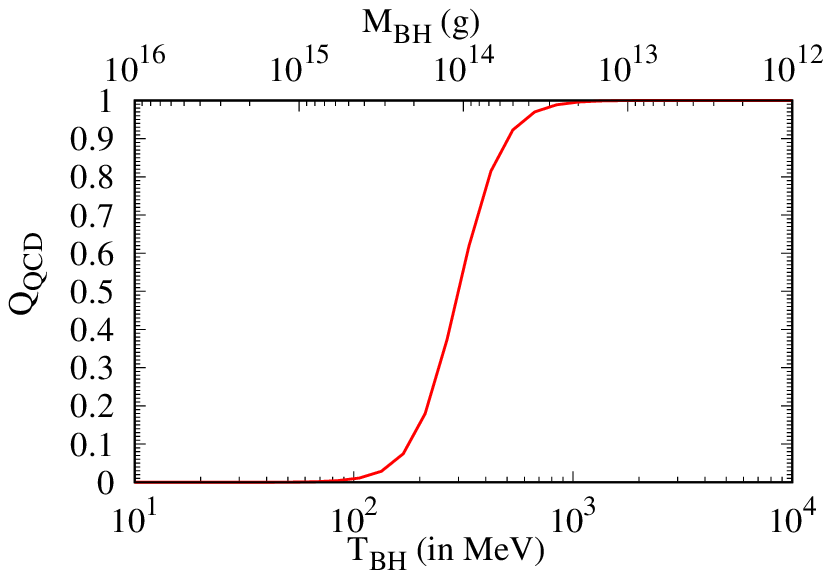}
    \caption{}
    \label{fig:qqcd}
  \end{subfigure}
  \caption{The parameters (a)$f(M_{\rm BH})$, (b) $Q_{\rm {QCD}}(T_{\rm BH})$
    defined in Eq. \ref{particle fraction} as a function of black hole mass and temperature.}
  \end{figure}

     We consider Schwarzschild black holes in this work. Since for a
     rotating black hole, more than 50 percent of the black hole energy is emitted when the black hole has already lost most of its spin \cite{P19761}, this is a good approximation. The spectrum of the emitted particles just outside the horizon of the black hole is given by a thermal distribution with Hawking temperature (with $c=\hbar=k_B$=1, where $c$ is the speed of light, $\hbar$=$\frac{h}{2\pi}$, $h$ is the Planck constant, $k_B$ is the Boltzmann constant) \cite{H1974,CKSY2010},
     \begin{equation}
     T_{\rm BH}=\frac{1}{8\pi GM_{\rm BH}}=1.06M^{-1}_{\mathrm{10}} \, \rm{TeV},
\end{equation}
where  $M_{10}=M_{\rm BH}/10^{10}g$ is the instantaneous mass of black hole  in units of $10^{10}$ g. 
Since, the emitted particles initially propagate in the gravitational
potential well of the black hole, the actual spectrum of emitted particles,
far from the black hole, deviates from a pure thermal spectrum and this
deviation is captured by the absorption coefficient $\Gamma_s(E,M_{\rm
  BH},m)$ \cite{P1976,MW1990}, where $s$ is the spin of the emitted
particle, $E$ is the energy of the emitted particle, $M_{\rm BH}$ is the
mass of the black hole, and $m$ is the mass of the emitted particle. The
effective potential as seen by a propagating particle outside the black
hole is a function of black hole mass,  particle mass and particle spin \cite{RW1957}. With this correction, the emitted spectrum of black hole in the energy interval $E$ and $E+dE$ is given by,
\begin{equation}
\frac{dN}{dt}=\frac{\Gamma_s(E,M_{\rm BH},m)}{2\pi}\frac{1}{e^{E/T_{\rm BH}}-(-1)^{2s}}dE.
\end{equation}
At high energy ($E>>T_{\rm BH}$), the absorption coefficients become
independent of particle mass and spin and approach the thermal limit
$\Gamma_s(E/T_{\rm BH}>>1)=27G^2M^2$. The evaporation of the black holes
changes their temperature which in turn changes the spectrum of the emitted particles. Therefore, for accurate calculation of energy cascades from black hole evaporation, we have to keep track of the black hole temperature or mass as it evaporates. The mass loss rate from evaporating black holes can be written as \cite{MW1991}, 
\begin{equation}
\frac{dM_{\rm BH}}{dt}=-5.34\times 10^{25} \newline
\frac{f(M_{\rm BH})}{M^2_{\rm BH}}\,{\rm  g s^{-1}}.
\label{evolution}
\end{equation} 
where $f(M_{\rm BH})$ is approximately given by,
\begin{multline}
f(M_{\rm BH})=1.0+0.569[\exp(-\frac{M_{\rm BH}}{x_{{\rm bh},s=1/2} M^T_{\rm e}})+\exp(-\frac{M_{\rm BH}}{x_{{\rm bh},s=1/2} M^T_{\mu}})+
3\exp(-\frac{M_{\rm BH}}{x_{{\rm bh},s=1/2} M^T_{u}})Q_{\rm{QCD}} \\
+3\exp(-\frac{M_{\rm BH}}{x_{{\rm bh},s=1/2} M^d_{d}})Q_{\rm{QCD}} +3\exp(-\frac{M_{\rm BH}}{x_{{\rm bh},s=1/2} M^T_{s}})Q_{\rm{QCD}}+ 
3\exp(-\frac{M_{\rm BH}}{x_{{\rm bh},s=1/2} M^T_{c}})Q_{\rm{QCD}}\\ +\exp(-\frac{M_{\rm BH}}{x_{{\rm bh},s=1/2} M^T_{\tau}})Q_{\rm{QCD}}
+3\exp(-\frac{M_{\rm BH}}{x_{{\rm bh},s=1/2} M^T_{b}})Q_{\rm{QCD}}]\\ +0.963\exp(-\frac{M_{\rm BH}}{x_{{\rm bh},s=1} M^T_{g}})Q_{\rm{QCD}}
+0.267\exp(-\frac{M_{\rm BH}}{x_{{\rm bh},s=0} M^T_{\rm \pi^0}})(1-Q_{\rm{QCD}})\\
+2.0\times 0.267[\exp(-\frac{M_{\rm BH}}{x_{{\rm bh},s=0} M^T_{\rm \pi^+}})(1-Q_{\rm{QCD}})],
\label{particle fraction}
\end{multline}
with
%+0.267[\exp(-\frac{M_{\rm BH}}{x_{{\rm bh},s=0} M^T_{\rm \pi^0}+2.0*0.267[\exp(-\frac{M_{\rm BH}}{x_{{\rm bh},s=0} M^T_{\rm \pi^+}}]
\begin{equation}
M^T_i=\frac{1}{8\pi Gm_i},
\end{equation}
 where $M^T_i$ is the mass of the black hole whose temperature is equal
 to the  mass $m_i$ of the standard model particle, where $i\in
 {(e,\mu,\pi^0,\pi^+,u,d,s,c,\tau,b,g)}$ for electron, muon, neutral and charged pions, up quark,
 down quark, strange quark, charm quark, tau, bottom quark and
 gluon  respectively, $x_{\rm bh}=\frac{E}{T_{\rm BH}}$, and $x_{{\rm bh},s}$ is
 the location of the peak of the emitted instantaneous power spectrum,
 $\frac{\gamma_s(x_{\rm bh})x^3_{\rm BH}}{e^{x_{\rm
       bh}}-(-1)^{2s}}$ with $\gamma_s(x_{\rm bh})=
 \Gamma_s(E)/27G^2M^2$. The value of $x_{{\rm bh},s}$ for s=0,1/2, and 1 is
 2.66,4.53, and 6.04 respectively \cite{MW1991}. We use the updated list of mass of particles from \cite{PDG2018}. The effect of QCD phase transition
 is captured in $Q_{\rm{QCD}}$ with
 $Q_{\rm{QCD}}=[1+\exp(\frac{-\log_{10}(T_{\rm
     BH}/T_{\rm{QCD}})}{\sigma})]$ \cite{SKLP2018,LSHLC2019},
 $T_{\rm{QCD}}$=300 MeV, and $\sigma$=0.1. We assume that above
 300 MeV, the kinematical mass of up and down quarks
 \cite{PYTHIA2006,PYTHIA2015}, quarks are freely emitted and below this energy only pions are produced as shown in Fig. \ref{fig:qqcd}.  
   \begin{figure}
\centering
\includegraphics[scale=1.0]{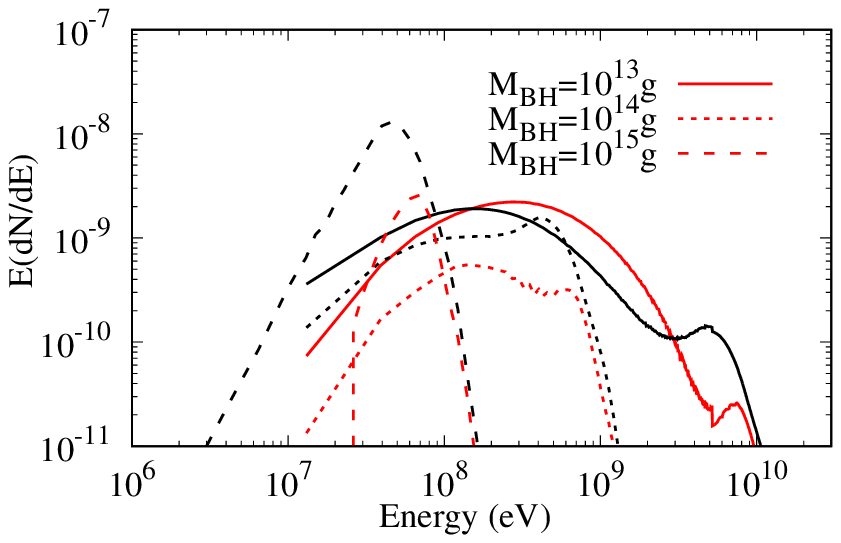}
\caption{Instantaneous spectrum of electron-positron pairs (black) and
  photons (red) at redshift $z=10000$ for different black hole mass. We
  have plotted the number of particles per unit log energy (EdN/dE).}
\label{fig:pbhspectrum}
\end{figure}  
   For the black holes with mass $M_{\rm BH}\gtrsim
  10^{13}$g, the contribution of W and Z bosons, higgs and top
    quark is negligible (Fig. \ref{fig:fmbh}) and therefore, is
  neglected. Freely emitted quarks, gluons, pions, muons and tau will
  hadronize or decay to produce secondary photons, electrons and
  positrons. To calculate the secondary photon, electron and positron
  spectrum, we have used $\rm{PYTHIA}$ 8.2 \cite{PYTHIA2006,PYTHIA2015}. We
  use the script available with $\rm{PYTHIA}$ 8.2 for
  hadronization of quarks and gluons and decay of taus, muons and any
  unstable hadrons produced during hadronization. The output obtained for a
  monochromatic particle (i.e. quark or muon injection etc.) is tabulated
  as spectra of stable standard model particles i.e. electrons, positrons,
  photons, neutrinos and stable hadrons (proton and deuterium). Any
  injection spectrum which is non-monochromatic can be thought of as a
  superposition of monochromatic energy injection and the tabulated results
  for monochromatic energies is used to obtain the final particle spectrum.
  In Fig. \ref{fig:pbhspectrum}, we plot the instantaneous spectra of
  emitted electron-positron pairs and photons at $z$=10000, when the black
  holes are almost intact. The temperature of black hole with mass
  $10^{13}$g is $\sim$ 1 GeV. The quarks and gluons produce secondary
  photons and electron-positron pairs which dominate the output
  spectrum. The peaks at high energy end is due to primary emission of
  photons and electron-positron pairs. For black hole mass $\gtrsim
  10^{14}$g, primary emission of photons, electrons and positrons dominate
  the output spectrum as quark, gluon, pion, and muon channels shut down. 
  \section{Energy deposition from black hole evaporation}
  Energy injection rate from evaporating black holes is given by,
  \begin{equation}
  \frac{dE_{\rm{inj}}}{dt}=f_{\rm{BH}}\rho_c c^2 (1+z)^3\frac{{dM_{\rm{BH}}/dt}}{M_{\rm{BH},0}}, 
  \end{equation}   
  where $f_{\rm{BH}}=\frac{\rho_{\rm{BH}}}{\rho_c}$ is the fraction PBH
  energy density ($\rho_{\rm{BH}}$) w.r.t to the stable cold dark matter energy density ($\rho_c$),  $M_{\rm{BH},0}$ is the mass of the black hole before evaporation. The energy injected in a time interval $\Delta t$ (in a redshift step $\Delta z$) is given by,
  $\Delta E_{\rm{inj}}= \frac{dE_{\rm{inj}}}{dt} \Delta t,$
  where $\Delta t=\frac{|\Delta z|}{(1+z)H(z)}$, and $H(z)$ is the
    Hubble rate. The deposited
energy fraction at a redshift $z$ is defined as,
  \begin{equation}
  f_{\rm{dep}}(z)=\frac{\Delta E_{\rm{dep}}/\Delta t}{\Delta E_{\rm{inj}}/\Delta t},
  \end{equation}
  where $\Delta E_{\rm{dep}}$ is the energy deposited during redshift interval between $z$ and $z-\Delta z$.
  The fraction of energy going into ionization of  neutral hydrogen  is defined as,
  \begin{equation}
  f_{\rm{H,ion}}(z)=\frac{\Delta E_{\rm{H,ion}}/\Delta t}{\Delta E_{\rm{inj}}/\Delta t},
  \end{equation}
where $\Delta E_{\rm{H,ion}} \equiv \Delta N_{\rm{H,ion}}\times 13.6~{\rm
  eV}$ and  $\Delta N_{\rm{H,ion}}$ is the number of  ionizations from
energy injection.
  \par
  \hspace{1cm}
   \begin{figure}
\centering
\includegraphics[scale=1.0]{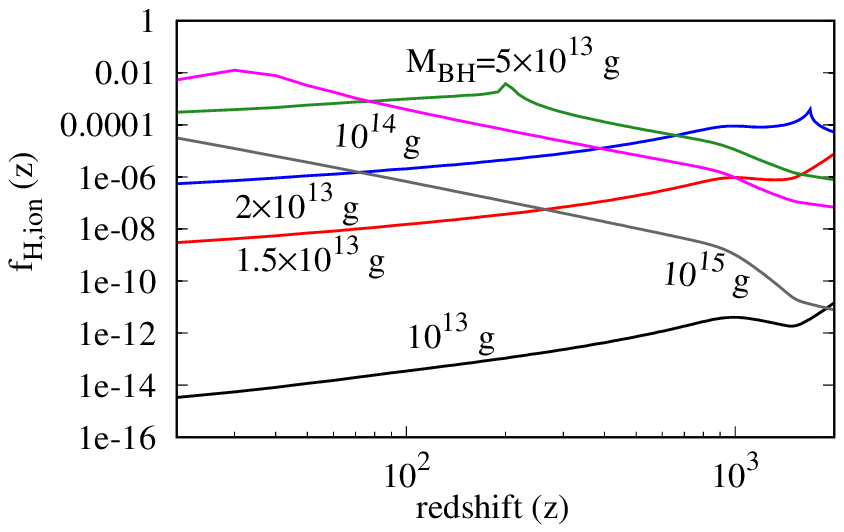}
\caption{Fraction of injected energy going to hydrogen ionization as a function
  of redshift for PBHs of different masses.}
\label{fig:Hdepfrac}
\end{figure}  
The energy injected by PBH is deposited in the background baryonic gas as
heating, excitation of neutral hydrogen and helium, and
  ionization of neutral hydrogen, neutral helium,  and singly ionized
helium. Low energy photons with energy less than the Lyman-alpha threshold
of hydrogen are not deposited to the baryonic gas and they have a
  completely 
  negligible 
  effect on the  recombination process. However, we still tag them as deposited energy for the purpose of accounting. These photons must be followed separately if we want to calculate their contribution to the CMB spectral distortions. Energy deposited at a particular redshift includes contribution from energy injection at that redshift and earlier redshifts.
In Fig. \ref{fig:Hdepfrac}, we plot the fraction of evaporating black
holes' energy going into hydrogen ionization as a function of redshift for
different values of PBH mass. To calculate the energy deposition fractions, 
standard recombination history \cite{Zks1969,Peebles1969} is assumed for
all cases. This is a good approximation since the modification to
the recombination history due to energy injection is small, and
therefore does not change the energy deposition fraction significantly. We
have used the \textbf{Recfast++} module \cite{Seager:1999bc,Chluba2010} of
\textbf{CosmoRec} code
\cite{CS2006,SH2008,RCS2008,Chluba2010,GH2010,CVD2010,AH2010,Hh2011} to
solve for the  recombination history with energy injection due to PBHs. We
have neglected excitations as they do not make a significant difference
\cite{AK20192}. A detailed discussion on the effect of the choice
of   recombination code on the CMB anisotropy power spectrum can be found in \cite{AK20192}.\par
\hspace{1cm}
 For the black holes evaporating before the recombination epoch
 ($M_{\rm{BH,0}}\lesssim 10^{13}$ g), a fraction of energy survives
 and is not deposited until after the recombination epoch, and thus can
 affect the CMB anisotropy power spectrum. The black holes with
 mass $M_{\rm{BH,0}}>10^{14}$g are mostly intact until
 reionization. The cuspy feature  around redshift 200 in Fig. \ref{fig:Hdepfrac} 
   corresponds approximately to the redshift at which the black hole with mass $5\times 10^{13}$g completely
   evaporates  and thus injects most of its mass-energy into the plasma
   i.e. the redshift corresponding to the lifetime of the black hole. As
 can be seen from Eq. \ref{evolution}, the rate of evaporation of black
 hole increases as the black hole gradually evaporates. Therefore, most of the black hole's energy is evaporated in a small redshift range \cite{AK20193}. This substantially increases the energy deposition at the redshift where most of the black hole evaporates giving rise to an increase in the energy deposition fraction. Once the black hole has evaporated away, there is no more energy injection and the deposited energy from energy injected at earlier times gradually decreases with time. For black hole with  mass heavier than $5\times 10^{13}$g, the black holes are still intact. Since, there is  tiny amount of energy injection from black holes, the fraction of energy deposited with respect to the unevaporated black hole mass is small. The deposition fraction of these heavier black holes increases with decreasing redshifts as black hole evaporation proceeds. For mass smaller than $5\times 10^{13}$ g, lesser and lesser energy are deposited with decreasing redshifts as most of the energy has been deposited at higher redshifts. 
 \section{Planck CMB anisotropy, BBN and CMB spectral distortions constraints on PBHs}
       \begin{figure}
\centering
\includegraphics[scale=1.5]{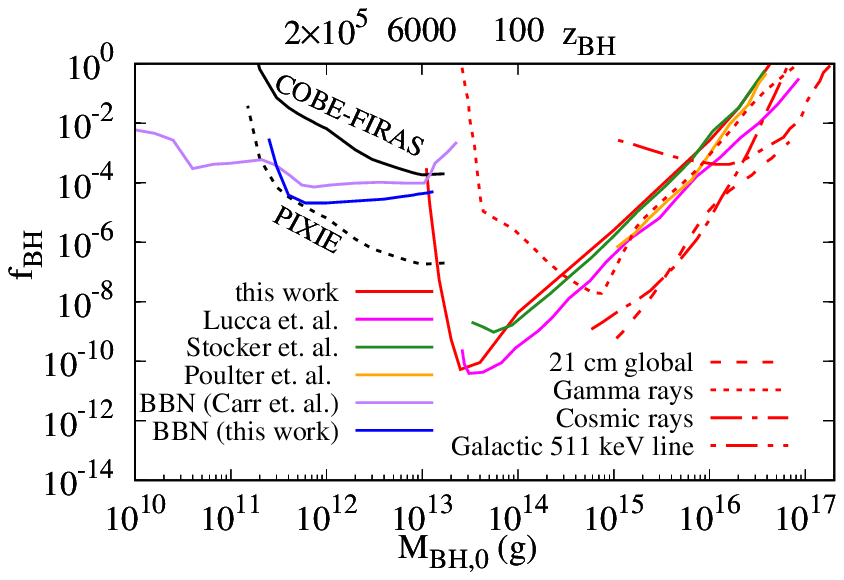}
\caption{2-$\sigma$ constraints (upper limits) on the abundance
  of PBHs as a function of mass derived in this work using CMB anisotropies
  and abundance of light elements. The evaporation redshift ($z_{\rm BH}$)
  corresponds to the redshift at which the mass of the black hole
    of a given mass reduces by a factor of e(=2.718). Comparison with
  previous works of Stocker et. al. \cite{SKLP2018}, Poulter
  et. al. \cite{PAHWW2019}, and  Lucca et. al. \cite{LSHLC2019} is
shown. We also show spectral distortion constraints using  COBE-FIRAS data
\cite{F1996,fm2002} derived in  \cite{AK20193} as well as  projection from
a future PIXIE-like experiment assuming a factor of $1000$ improvement over
COBE-FIRAS. We also show BBN constraints  from Fig. 9 of
\cite{LSHLC2019} which were adapted from \cite{KY2000,CKSY2010},
constraints from global 21 cm signal \cite{CDGMS2018}, gamma rays (taken
from \cite{AAS2019}, non-spinning case), cosmic rays \cite{BC2019} (model A
with background), and 511 keV gamma ray line from galactic center (taken
from \cite{DLR2019}) for comparison.}   
\label{fig:pbhconst}
\end{figure}  
\begin{figure}
\centering
\includegraphics[scale=1.5]{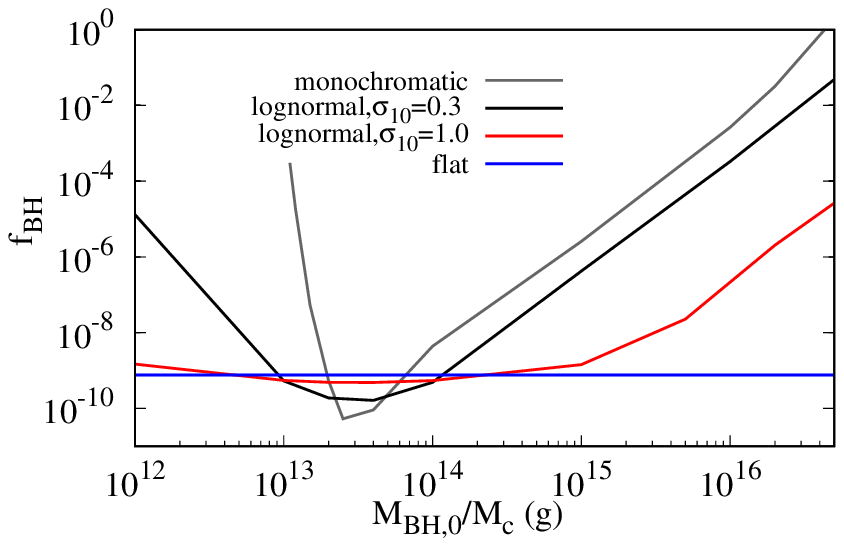}
\caption{Constraints on black holes with extended mass
  distribution. We consider lognormal distribution with two different
  variances and a flat distribution with lower and upper limits for mass
  being  $10^{12}$g and $5\times 10^{16}$g respectively. The x-axis denotes $M_{\rm{BH,0}}$ for monochromatic and $M_c$ for lognormal mass distribution in Eq. \ref{lognormal}.}  
\label{fig:extspecrumconst}
\end{figure} 
     \begin{figure}
\centering
\includegraphics[scale=1.5]{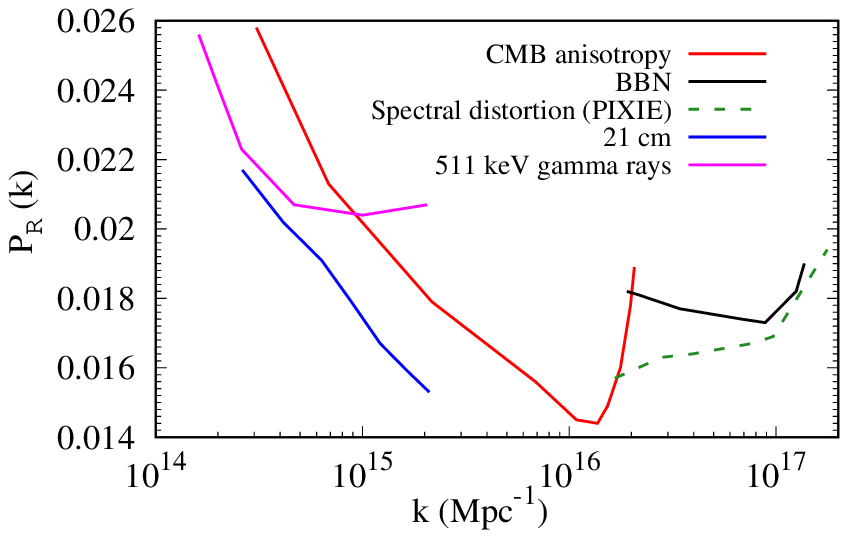}
\caption{Constraints on primordial curvature power spectrum as a function of comoving wave number from CMB anisotropies, BBN calculations in this work and PIXIE \cite{Pixie2011} projection from \cite{AK20193}. Also shown are 21 cm and 511 keV gamma rays constraints from Fig. \ref{fig:pbhconst}, which are the strongest constraints in the mass range $10^{15}$g -$10^{17}$g. }  
\label{fig:powerspecrumconst}
\end{figure}  
We show the main results of this paper in Fig. \ref{fig:pbhconst}. We do a
Markov chain Monte carlo (MCMC) analysis with the publicly available code
\textbf{COSMOMC} \cite{LB2002} and use \textbf{Recfast++} module of \textbf{CosmoRec}
\cite{Chluba2010} to solve for the  recombination history. We fit for
 $f_{\rm  BH}$ along with the 6 standard cosmological parameters
\cite{Pl2018} using Planck2018 PlikTT,TE,EE and low E likelihood
\cite{Pl2019}. We assume monochromatic mass function and perform the fit
for different initial black hole mass $M_{\rm BH,0}$ and obtain constraints
on $f_{\rm BH}$ as a function of $M_{\rm BH,0}$. We show 2-$\sigma$ upper limits in Fig. 4. For abundance
of light elements from BBN, we find that the abundance of helium-3 ($^3\rm{He}$)
gives the strongest  constraints. The initial abundance of $^3\rm{He}$
before black hole evaporation is taken to be the theoretical 2-$\sigma$
lower limit prediction from BBN for Planck cosmological
parameters. Constraints are obtained by allowing the maximum abundance,
after creating extra $^3\rm{He}$ from $^4\rm{He}$ photo-dissociation, to be
at the 2-$\sigma$ observed upper limit. We also show the CMB spectral
distortion constraints obtained in \cite{AK20193} from Cosmic
Background Explorer-Far Infrared Absolute Spectrophotometer (COBE-FIRAS)  data and projections for the Primordial Inflation
  Explorer (PIXIE) or similar 
mission assuming a factor of 1000 improvement over COBE-FIRAS.   \par
\hspace{1cm}
Our results show that the Planck CMB anisotropy measurements put stronger
constraints compared to the BBN and CMB spectral distortions from
COBE-FIRAS \cite{F1996} on the abundance of primordial black
holes with mass $M_{\rm BH,0}\geq 1.1\times 10^{13}$g. This limit is slightly
lower compared to the mass of $3\times 10^{13}$g upto which CMB anisotropy
constraints were calculated in  \cite{SKLP2018,LSHLC2019}. A fraction of
high energy photons injected before the recombination epoch can survive
 until the end of the recombination epoch and deposit their energy once
the universe becomes neutral.  The much high precision with which the CMB
anisotropies are measured and the sensitivity of the CMB anisotropy power
spectrum to small changes in the recombination history means that the CMB
anisotropy power spectrum can  provide competitive CMB anistropy
constraints compared to BBN and spectral distortions even for
pre-recombination energy injection.
Our results agree with the results of Lucca et. al. \cite{LSHLC2019} for
black holes evaporating at $z\sim$ 1000. Lucca
et. al. use on-the-spot approximation for energy deposition which is a reasonable approximation for energy injections at $z\sim$ 1000 but this approximation will over-estimate the deposition
efficiency at lower redshifts. Therefore, we expect our constraints without
on-the-spot approximation to be weaker for  black holes evaporating at
redshifts $z<$1000. Our results agree with the result of Stocker
et. al. \cite{SKLP2018} for black holes with initial temperatures less than
the QCD phase transition. The authors in \cite{SKLP2018} use the results of
Slatyer et. al. \cite{Slatyer20162} to calculate the energy deposition
fractions. However, there may be  possible numerical errors in taking into
account QCD phase transition in \cite{SKLP2018} (see discussions in
\cite{LSHLC2019}).  This might explain the difference between  our results and \cite{SKLP2018} for PBH mass smaller than $10^{14}$g, precisely at the point when the black hole temperature is of the order $\sim$100 MeV. We have reasonable agreement with the results of Poulter et. al. \cite{PAHWW2019} who also use the results of \cite{Slatyer20162} without on-the-spot approximation. For the mass range considered in \cite{PAHWW2019}, temperature of black hole is below the QCD phase transition. However, the authors have made approximations such as considering the universe to be completely matter dominated below $z<3000$ which would have affected their quantitative results. \par
\hspace{1cm}
We also show our BBN constraints with full evolution of electromagnetic
cascades in an expanding universe. For comparison, we also plot BBN
constraints from Fig. 9 of \cite{LSHLC2019} which is adapted from
\cite{CKSY2010,KY2000}. The BBN constraints in \cite{CKSY2010,KY2000} are
derived taking into account both electromagnetic and hadronic interactions
of emitted particles (electrons, positrons, photons, nucleons and
antinucleons) with background electrons, photons and
hadrons. Since we ignore the hadronic interactions with primordial
  nuclei which would
  also result in their destruction, our results would be expected to be
  slightly weaker or conservative compared to  \cite{CKSY2010,KY2000}.
Nevertheless, we would like to point out that with the current constraints
on $^3\rm{He}$, significantly stronger constraints than the constraints in
\cite{CKSY2010} are possible. The authors in \cite{CKSY2010} consider
2-$\sigma$ upper bound on ratio of deuterium $^2\rm{H}$ to hydrogen ($\rm{H}$) abundance to be,
$\frac{n_{^2\rm{H}}}{n_{\rm{H}}}=5.16\times 10^{-5}$ and for ratio of 
$^3\rm{He}$  to deuterium to  be 
$\frac{n_{^3\rm{He}}}{n_{^2\rm{H}}}=1.37$,  where $n_{^3\rm{He}}$,
$n_{^2\rm{H}}$ and $n_{\rm{H}}$ are the observed helium-3, deuterium and
hydrogen number density respectively. Our 2-$\sigma$ upper limit on
$^3\rm{He}$ abundance is $\frac{^3\rm{He}}{\rm{H}}=1.5\times 10^{-5}$
\cite{BRB2002}. With the new upper limits from observations but neglecting
energy going into stable nucleons, we conservatively get BBN constraints which are stronger than the constraints in \cite{CKSY2010} by upto a factor of 4-5. The BBN constraints in \cite{CKSY2010} for black hole mass below $\lesssim 4\times 10^{11}$g are due to lithium which we do not consider in this work. \par
\hspace{1cm} 
We also show the CMB spectral distortion constraints from COBE-FIRAS
\cite{F1996} derived in \cite{AK20193} and projections for PIXIE
\cite{Pixie2011} assuming a factor of 1000 improvement over COBE-FIRAS. The
spectral distortion constraints are derived with full electromagnetic
cascade evolution which can be substantially different from thermal ($y,i$
or $\mu$) distortions due to contributions from non-thermal relativistic
particles. For primordial black holes with mass $\approx
10^{13}$g, the actual constraints, taking into account  the actual shape of
non-thermal distortions, are relaxed by  a factor of 2 compared to the
constraints using the yim approximation. \par
\hspace{1cm}
We also show, for comparison, the  constraints from 21 cm line observations, gamma rays, cosmic rays, and 511 keV galactic gamma rays  in Fig. \ref{fig:pbhconst}.
Energy injection from primordial black hole evaporation can heat
and ionize the intergalactic medium and modify the
  global 21 cm spin temperature evolution. Requiring that the 21
cm absorption signal during the reionization epoch does not get
wiped out from electromagnetic energy injection, we can put constraints on
evaporating black holes \cite{CDGMS2018}. However, these constraints have
to be re-derived considering EDGES \cite{B2018} result which require new
physics or consideration of astrophysical uncertainties. Primordial black holes evaporating today ($M_{\rm{BH,0}}\sim 10^{15}$g) emit photons and electron-positron pairs dominantly. Using the diffuse gamma ray background \cite{AAS2019}, cosmic ray $e^-e^+$ \cite{BC2019} and 511 keV gamma rays from annihilation of $e^-e^+$ at galactic center \cite{DLR2019}, we can put constraints on abundance of evaporating black holes in the mass range $M_{\rm{BH,0}}\gtrsim 10^{15}$g. \par
\hspace{1cm}
  In Fig. \ref{fig:extspecrumconst}, we show the constraints on black holes with extended mass functions. We use the prescription of \cite{CRTVV2017} to convert monochromatic mass constraints to extended mass constraints. We consider lognormal distribution with mass function,
  \begin{equation}
  \psi(M_{\rm BH,0})=\frac{f_{BH}}{\sqrt{2\pi}\sigma M_{\rm
      BH,0}}\exp\left(-\frac{(\ln(M_{\rm BH,0}/M_c)^2}{2\sigma^2}\right),
  \label{lognormal}
  \end{equation}
  where $M_c$ is the center of mass distribution and $\sigma$ is the
  variance. We can convert the base in the log from e to 10 to suit our
  numbers \cite{PAHWW2019}. In that case
  $\sigma=\sigma_{10}\mathrm{ln(10)}$. The constraints for extended mass
  spectrum with center $M_c$ are in general stronger compared to
  monochromatic spectrum due to contribution of stronger constraints from
  monochromatic mass with $M_{\rm BH,0}$ different from $M_c$.  In the opposite case,
  when the monochromatic constraints with $M_{\rm BH,0}=M_c$ are the
  strongest, the presence of other black holes with different mass in the
  extended mass function make the constraints for extended mass function
  weaker.  The black holes with different masses have different energy
  injection histories and the effect of
  black hole evaporation on the CMB for different masses will not be
  strictly additive. These constraints are therefore only approximate but
  sufficient to give a qualitative idea about what to expect for extended
  mass functions.

\section{Constraints on primordial power spectrum and  40 e-folds of inflation}
 Since PBH are formed from fluctuations of density perturbations in
 the early Universe, we can put constraints on the
   initial  power spectrum from the allowed abundance of PBH
 \cite{C1975,Ch1975,CGL1994,INN1994,JGM2009,SKK2019}. In
 Fig. \ref{fig:powerspecrumconst}, we show constraints on primordial
 curvature power spectrum $P_R(k)$ as a function of comoving wave number
 ($k$) from CMB anisotropies and BBN calculations, done in this work, as
 well as PIXIE forecasts which can be in principle be stronger compared to
 the current BBN constraints. We also show 21 cm and 511 keV gamma rays
 constraints, which are the strongest constraints in the  low $k$
 part of parameter space. We follow the procedure of \cite{SKK2019} to
 translate the constraints on $f_{\rm BH}$ to the constraints on  $P_R(k)$. Mass of PBH formed in a particular epoch is given by, 
\begin{equation}
M_{\rm BH,0}=\frac{4\pi}{3}\gamma \rho H^{-3},\label{Eq:mbh}
\end{equation}
where $\gamma=0.2$ \cite{C1975,CKSY2010}, $\rho$ is the average total energy density of the Universe and $1/H$ is the horizon size at that epoch.  
 In Press-Schechter theory, PBH form if the smoothed density perturbation
 is above a certain threshold $\delta_c$, which was derived to be 0.42 in a
 radiation dominated universe in \cite{HYK2013,EGS2020}. The density
 field on smoothing scale $R$, $\delta(R)$, is assumed to have a gaussian distribution,  
 \begin{equation}
 P(\delta(R))=\frac{1}{\sqrt(2\pi)\sigma(R)}\exp\left(-\frac{\delta^2(R)}{2\sigma^2(R)}\right),
 \end{equation}
  where $\sigma^2(R)=\int_{0}^{\infty}W^2(kR)P_\delta (k)
  \frac{dk}{k}$,  $R=1/(aH)$,   $a$ is the scale factor, $W(kR)$ is a window
  function to smoothen the density perturbations which we assume to be
  Gaussian. The relative abundance of PBH is obtained by integral of the
  probability distribution of density perturbation with density fluctuations greater than $\delta_c$, i.e.
\begin{equation}
\beta=2  \int_{\delta_c}^{1} d\delta(R)P(\delta(R))\approx \erfc\left(\frac{\delta_c}{\sqrt 2\sigma(R)}\right).
\label{beta}
\end{equation}
We can obtain $\beta$ from $f_{\rm{BH}}$ by using the relation \cite{SKK2019},
\begin{equation}
\beta(M_{\rm{BH,0}})=4.0\times 10^{-9}\left(\frac{g^i_*}{10.75}\right)^{1/4}\left(\frac{M_{\rm{BH,0}}}{2\times 10^{33} g}\right)^{1/2}f_{\rm{BH}},
\end{equation}
where $g^i_*$ is the total radiative degrees of freedom at the
formation epoch of PBH. The average energy of the standard model
  particles or the temperature of the Universe at the formation epoch of
black holes considered in this work is in the range of
$10^{17}$eV-$10^{20}$eV. Since, this  energy scale  is much
higher than the mass scale of all standard model particles, we take
$g^i_*$ to be 106 using the standard model of particle physics.
 The relation to convert density perturbation to curvature perturbation is given by \cite{GLMS2004},
  \begin{equation}
  \delta(k,t)=\frac{2\left(1+3w\right)}{(5+3w)}(k/aH)^2 \mathcal{R},
  \end{equation}  
  with $w=1/3$, where $w$ is the equation of state in the radiation dominated era. Power spectrum of density fluctuations can then be related to power spectrum of curvature perturbation as,
  \begin{equation}
 P_{\delta}(k,t)=\left(\frac{2(1+3w)}{(5+3w)}\right)^2(k/aH)^4 P_{\mathcal{R}}(k),
\label{powerspectrum}  
  \end{equation}
  We assume slow-roll like inflationary power spectrum
  $P_\mathcal{R}(k)=P_\mathcal{R}(k_0)(\frac{k}{k_0})^{n_s-1}$ with $n_s
  \sim 1$. Obtaining $\sigma(R)$ from Eq. \ref{beta} and substituting
  Eq. \ref{powerspectrum} in the expression of $\sigma(R)$, we calculate
  $P_\mathcal{R}(k_0)$ for a given $f_{\rm BH}(M_{\rm BH,0})$ with $k_0=aH$,
  where $a$ and $H$ are the scale factor and  Hubble parameter at the time of black hole formation related
  to $M_{\rm BH,0}$ by Eq. \ref{Eq:mbh}.  
 In Fig. \ref{fig:powerspecrumconst}, we provide constraints on amplitude
 of the primordial power spectrum from constraints on abundance of PBH
 obtained in this work as well as 21 cm and 511 keV galactic gamma rays constraints
     which are stronger in appropriate mass range. Together, these probes put
 an upper bound on the amplitude of the primordial power spectrum on the
 scales of $10^{14}$ Mpc$^{-1}\lesssim k \lesssim 10^{17}$
 Mpc$^{-1}$. The perturbations on these scales were created
   towards the end of inflation, about
   $\sim 40$ e-folds after the modes observed in the CMB anisotropies and
   large scale structure went our of horizon during inflation. Therefore, combined
   with the CMB anisotropies, PBH provide a view of almost the entire
   inflationary history. In particular, the abundance of PBH are almost the
 only observational constraints on this epoch of inflation just before
 inflation ends. Constraints on the
abundance of the PBH in the mass range $10^{11}$g-$10^{17}$g nicely complement the constraints from heavier black holes ($k \lesssim 10^{14}$ Mpc$^{-1}$ \cite{SKK2019}) and CMB spectral distortions constraints from  acoustic damping on the scales of $ k \lesssim 10^{4}$ Mpc$^{-1}$ \cite{Ks2013}.
\section{Conclusions}
In this work, we provide CMB anisotropy power spectrum constraints and BBN
constraints on evaporating black holes with our own calculations of the evolution of the electromagnetic particle cascade
in an expanding universe. Our constraints are based on the
precise treatment of electromagnetic cascades in a consistent framework for
all three data sets relevant during and before recombination, namely, CMB
anisotropies, spectral distortions, and BBN abundances of elements. We have
shown that the CMB anisotropies can constrain black holes with slightly
lower mass compared to the previous calculations. We obtain good agreement
with previous literature in the regime where the approximations used in
those calculations are reasonable. We also provide BBN constraints with
more up-to-date observational values of helium-3 abundance. Even with a factor of 1000
improvement in future with PIXIE-like mission, CMB anisotropies will still
provide stronger constraints for $M_{\rm BH,0}\geq 10^{13}$g. Our work, in
particular, fills the small gap that existed  between the
BBN/spectral distortions constraints and CMB anisotropy constraints. Even
though the black holes considered in this paper would have been completely
evaporated by now, the limits on their existence in the early universe
provide important constraints on perturbations generated during the last
stages of inflation corresponding to the modes on scales $10^{14}$
Mpc$^{-1}\lesssim k \lesssim 10^{17}$ Mpc$^{-1}$,  $\sim
  40$ e-folds after the modes we observe in the CMB anisotropies and large
  scale structure of the Universe went out of horizon during inflation. 
 \section{Acknowledgements}
 We acknowledge the use of computational facilities of Department of
 Theoretical Physics at Tata Institute of Fundamental Research,
 Mumbai. This work was supported by Max Planck Partner Group for cosmology of Max Planck
 Institute for Astrophysics Garching at 
 Tata Institute of Fundamental Research funded by
 Max-Planck-Gesellschaft. This work was also supported by 
Science and Engineering Research Board (SERB) of Department of Science and
Technology, Government of India grant no. ECR/2015/000078. We acknowledge support of the Department of Atomic Energy, Government of India, under project no. 12-R\&D-TFR-5.02-0200.  
\bibliographystyle{unsrtads}
\bibliography{pbhconst1} 
\appendix
\section{Comparison of CMB constraints from Planck 2015 and 2018 data}
\label{tab:table}
\begin{table}[h!]
\fontsize{9}{9}\selectfont 
\caption{Comparison of constraints on abundance of black hole $f_{\rm{BH}}$
  from Planck 2015 \cite{Plik2015} and Planck 2018 \cite{Pl2019} data for
  monochromatic black hole mass function.}
  \begin{center}  
    \begin{tabular}{l|c|r|l|c|r|l} % <-- Alignments: 1st column left, 2nd middle and 3rd right, with vertical lines in between
      %\textbf{V} & \textbf{Value 2} & \textbf{Value 3}\\
     $M_{\rm{BH,0}}$ (g) & P2015 & P2018  \\
      \hline
    $1.1\times 10^{13}$ & $3.92\times 10^{-4}$ & $3.05\times 10^{-4}$  \\
    $1.2\times 10^{13}$ & $1.81\times 10^{-5}$ & $1.78\times 10^{-5}$  \\
    $1.5\times 10^{13}$ & $6.48\times 10^{-8}$ & $5.32\times 10^{-8}$  \\
    $2\times 10^{13}$ & $6.53\times 10^{-10}$ & $5.36\times 10^{-10}$  \\
    $2.5\times 10^{13}$ & $7.62\times 10^{-11}$ & $5.31\times 10^{-11}$  \\
    $4\times 10^{13}$ & $9.4\times 10^{-11}$ & $9.1\times 10^{-11}$  \\
    $ 10^{14}$ & $5\times 10^{-9}$ & $4.36\times 10^{-9}$  \\
    $ 10^{15}$ & $2.7\times 10^{-6}$ & $2.58\times 10^{-6}$  \\
    $ 10^{16}$ & $2.62\times 10^{-3}$ & $2.54\times 10^{-3}$  \\
    $2\times 10^{16}$ & $3.56\times 10^{-2}$ & $3.2\times 10^{-2}$  \\
    $5\times 10^{16}$ & $2.16$ & $1.94$  \\
   
    \end{tabular}
  \end{center}
   \end{table}
\end{document}